\documentstyle[12pt]{article}
\newcommand{\Od}{{\cal O}}

\newcommand{\lsim}   {\mathrel{\mathop{\kern 0pt \rlap
  {\raise.2ex\hbox{$<$}}}
  \lower.9ex\hbox{\kern-.190em $\sim$}}}
\newcommand{\gsim}   {\mathrel{\mathop{\kern 0pt \rlap
  {\raise.2ex\hbox{$>$}}}
  \lower.9ex\hbox{\kern-.190em $\sim$}}}

\def\gappeq{\mathrel{\rlap {\raise.5ex\hbox{$>$}}
{\lower.5ex\hbox{$\sim$}}}}

\def\lappeq{\mathrel{\rlap{\raise.5ex\hbox{$<$}}
{\lower.5ex\hbox{$\sim$}}}}

\begin{document}
\input epsf \renewcommand{\topfraction}{0.8}
\pagestyle{empty}
\begin{flushright}
{CERN-TH/2000-258}
\end{flushright}
\vspace*{5mm}
\begin{center}
\Large{\bf Primordial magnetic fields from metric perturbations} \\
\vspace*{1cm}
\large{\bf Antonio L. Maroto}\\
\vspace{0.3cm}
\normalsize
CERN Theory Division, \\CH-1211 Geneva 23, Switzerland, \\and \\
Dept. F\'{\i}sica Te\'orica, \\Universidad Complutense de Madrid, 
\\28040 Madrid, Spain\\
\vspace*{2cm}
{\bf ABSTRACT} \\ \end{center}
\vspace*{5mm}
\noindent
We study the amplification of electromagnetic vacuum fluctuations
induced by the evolution of scalar metric  perturbations at the end of
inflation. Such perturbations break the conformal invariance of
Maxwell
equations in  Friedmann-Robertson-Walker backgrounds and allow
the growth of magnetic
fields on super-Hubble scales. We  relate 
the strength of the fields generated by this mechanism with the
power spectrum of scalar perturbations and estimate the amplification
on galactic
scales for different values of the spectral index. Finally
we discuss the possible effects of finite conductivity during
reheating.
\vspace*{0.5cm}

\vfill\eject

\setcounter{page}{1}
\pagestyle{plain}

\section{Introduction}

The existence of cosmic magnetic fields with large coherence lengths
($>10$ kpc) and  typical strength of $10^{-6}$ G, still remains
an open problem in astrophysics \cite{Kronberg}.
 A partial explanation, widely
considered in the literature,  is based on
the amplification of seed fields by means of the so called
galactic dynamo mechanism. In this mechanism, the differential rotation of the
galaxy is able to transfer energy into the magnetic field, but
nevertheless
it still  requires a pre-exisiting field to
be amplified. The present bounds on the necessary seed fields
to comply with  observations are in the range $B_{seed}\gsim
10^{-17}-10^{-22}$ G ($h=0.65-0.5$) at decoupling time, 
coherent on a comoving scale
of  $\lambda_G \sim 10$ kpc, for a flat universe
without cosmological
constant. For a flat universe with nonvanishing cosmological constant,
the limits can be relaxed up to $B_{seed} \gsim 10^{-25}-10^{-30}$ G  
($h=0.65-0.5$)  
at decoupling for $\Omega_\Lambda=0.7$ and $\Omega_m=0.3$ \cite{Davis}.
The observations of micro-Gauss magnetic fields in two high-redshift objects
(see \cite{Kronberg,Davis} and references therein) 
could, 
if correct, impose
more stringent conditions on the seeds fields or even on 
the dynamo mechanism itself. 

The cosmological origin of the seed fields is one of the most
interesting
possibilities, although some other mechanisms at the astrophysical
level,  such as the
Biermann battery process, have also been considered
\cite{Biermann,Harrison}.
In the
cosmological
case, in which
 we will be mainly interested in this work, it is
natural to expect \cite{turner} that the same mechanism that
gave rise to the large-scale galactic structure, i.e. amplification
of quantum fluctuations during inflation,
was also responsible for the generation of the primordial magnetic
fields. However, it was soon noticed  \cite{turner} that
the gravitational amplification
 does not operate in the case of electromagnetic (EM)
fields. This is  because of
the conformal triviality of Maxwell
equations in Friedmann-Robertson-Walker (FRW) backgrounds, i.e.
conformally
invariant equations in a conformally flat space-time.  In order to
avoid
this difficulty, several production mechanisms have been proposed in which
Maxwell equations are modified in different ways. Thus for example,
the addition of
mass terms to the photon or higher-curvature terms in the Lagrangian
was  studied in \cite{turner}. The  contribution of the conformal
anomaly was included in \cite{Dolgov}. In the context of string
cosmology, the effects of a dynamical dilaton field were taken into
account
in \cite{giovan}. Other examples include  non-minimal
gravitational-electromagnetic coupling \cite{opher}, inflaton
coupling to EM Lagrangian \cite{Ratra}, spontaneous breaking
of Lorentz invariance \cite{Mota} or 
backreaction of minimally coupled charged scalars \cite{proko,calz,Gio}. Some
of
them are able to generate fields of the required strength to seed
the galactic dynamo or even to account for the observations
without further amplification.

In this paper we explore the alternative possibility, i.e. we avoid
conformal triviality by considering deviations from the FRW metric
(see \cite{Bassett} for a suggestion along these lines). 
This approach is rather natural since  we  know that galaxies
formed from  small metric inhomogeneities   present at large scales
and,
in addition, it  does not require any modification of Maxwell
electromagnetism. In the inflationary cosmology, metric perturbations
are generated when quantum fluctuations become super-Hubble sized and
thereafter  evolve as classical fluctuations, reentering the horizon
during radiation or matter dominated eras \cite{brand}. 
The same mechanism would operate on large-scale  EM fluctuations. However, 
if 
conformal
invariance is not broken, each positive or negative frequency EM mode will 
evolve independently, without mixing. This implies that photons cannot be 
created and
therefore magnetic fields are not amplified. However, in the presence of an 
inhomogeneous
background, we will show that the mode-mode coupling between EM and metric 
perturbations 
generates the mixing. This in turn will allow us to relate the strength of 
the magnetic
field created by this mechanism and the particular form of the metric 
perturbations
described by the corresponding power spectrum. 
Those photons produced in the 
inflation-radiation transition with very long wavelengths 
can be seen as static 
electric or magnetic fields. Because of 
the high conductivity of the Universe in the radiation era, the electric
components are rapidly damped whereas, thanks to magnetic flux conservation,
the magnetic fields will remain 
frozen in the plasma  and their
subsequent evolution will be trivial, $B a^2=$const
\cite{turner,Ratra}. The paper is organized as follows. In section 2
we obtain the Maxwell equations in the presence of an inhomogeneous background
and calculate the occupation number of the photons produced. 
In section 3 we apply
these results to calculate the corresponding magnetic field generated at
galactic scales. Section 4 is devoted to the analysis of the effects of
finite conductivity in those results and finally, section 5 includes the
main conclusions of the paper.

\section{Maxwell equations and photon production}

Although there are previous
works on the production of scalar 
and fermionic particles
in inhomogeneous backgrounds \cite{Frieman,noise}, in this paper
we will need to extend the analysis to the case of gauge fields.
Let us then consider Maxwell equations
\begin{eqnarray}
\nabla_\mu F^{\mu\nu}=0,
\label{Maxwell}
\end{eqnarray}
in a background metric that can be splitted as
$g_{\mu\nu}=g^{0}_{\mu\nu}+h_{\mu\nu}$, where
\begin{eqnarray}
g^{0}_{\mu\nu}dx^\mu dx^\nu=a^2 (\eta)(d\eta^2-\delta_{ij}dx^idx^j)
\end{eqnarray}
is the flat FRW metric in conformal time and
\begin{eqnarray}
h_{\mu\nu}dx^\mu dx^\nu=2a^2(\eta)\Phi (d\eta^2+\delta_{ij}dx^i dx^j)
\end{eqnarray}
is the most general form of the linearized  scalar metric perturbation in the
longitudinal gauge and where it has been assumed that the spatial
part of the energy-momentum tensor is diagonal, as indeed happens in
the inflationary or perfect fluid cosmologies \cite{brand}. In this expression
$\Phi(\eta,\vec x)$ is the gauge invariant gravitational potential.
The equation (\ref{Maxwell}) can be written as:
\begin{eqnarray}
\frac{1}{\sqrt{g}}\frac{\partial}{\partial
  x^\mu}\left(\sqrt{g}g^{\mu\alpha}
g^{\nu\beta}(\partial_\alpha A_\beta-\partial_\beta
  A_\alpha)\right)=0,
\end{eqnarray}
which leads in this background to the following linearized equations
\begin{eqnarray}
\frac{\partial}{\partial x^i}\left((1-2\Phi)(\partial_i A_0-\partial_0
  A_i)
\right)=0,
\label{zeroeq}
\end{eqnarray}
for $\nu=0$ and
\begin{eqnarray}
\frac{\partial}{\partial \eta}\left((1-2\Phi)(\partial_i A_0-\partial_0
  A_i)
\right)\nonumber \\
+\frac{\partial}{\partial x^j}\left((1+2\Phi)(\partial_j A_i-\partial_i
  A_j)
\right)=0,
\label{ieq}
\end{eqnarray}
for $\nu=i$. In addition, we will use the Coulomb gauge condition 
$\vec \nabla \cdot \vec A=0$.

In order to study the amplification of vacuum fluctuations, let us 
consider a particular
solution of the above equations that we will denote by 
$A_\mu^{\vec k,\lambda}(x)$ such
that asymptotically in the past it behaves as a positive frequency plane wave
with momentum $\vec k$ and polarization $\lambda$, i.e,
\begin{eqnarray}
A_\mu^{\vec k,\lambda}(x)
\stackrel{\eta\rightarrow -\infty}{\rightarrow} 
A_\mu^{(0)\vec k,\lambda}(x)=\frac{1}{\sqrt{2kV}}
\epsilon_\mu(\vec k,\lambda)e^{i(\vec k\vec x-k\eta)}
\label{ord0}
\end{eqnarray}
where $k^2=\vec k^2$.  For the two 
physical polarization states we have, 
$\vec\epsilon(\vec k,\lambda)\cdot \vec k=0$ and $\epsilon_0(\vec k,\lambda)=0$.
We will work in a finite box with comoving volume $V$ and we will 
take the continuum
limit at the end of the calculation. We are assuming that metric 
perturbations vanish before inflation starts, so that we can define
an appropriate initial conformal vacuum state. Because of the presence of the 
inhomogeneous
background, in the asymptotic future, this solution will behave as a linear
superposition of positive and negative frequency modes with different momenta 
and
different polarizations, i.e.,
\begin{eqnarray}
A_\mu^{\vec k,\lambda}(x)\stackrel{\eta\rightarrow \infty}{\rightarrow} 
\sum_{\lambda'}\sum_{q}\left(\alpha_{kq \lambda\lambda'}
\frac{\epsilon_\mu(\vec q,\lambda')}{\sqrt{2qV}}
e^{i(\vec q\vec x-q\eta)}+\beta_{kq \lambda\lambda'}
\frac{\epsilon_\mu^*(\vec q,\lambda')}{\sqrt{2qV}}
e^{-i(\vec q\vec x-q\eta)}\right)
\label{asin}
\end{eqnarray}
It is possible to obtain an expression for the Bogolyubov coefficients 
$\alpha_{kq \lambda\lambda'}$ and $\beta_{kq \lambda\lambda'}$ to first order in the
metric perturbations. With that purpose, we look for solutions of the
equations of motion in the form:
\begin{eqnarray}
A_\mu^{\vec k,\lambda}(x)=A_\mu^{(0)\vec k,\lambda}(x)+A_\mu^{(1)\vec k,\lambda}(x)+ ... 
\label{expansion}
\end{eqnarray}
where $A_\mu^{(0)\vec k,\lambda}(x)$ is the solution in the absence of perturbations
given by (\ref{ord0}). 
Introducing this expansion in (\ref{zeroeq}) and Fourier transforming, 
we obtain for the
temporal component of the EM field to first order in the perturbations:
\begin{eqnarray}
A_0^{(1)\vec k,\lambda}(\vec q,\eta)=-\sqrt{\frac{2k}{V}}\frac{\vec q \cdot 
\vec \epsilon(\vec k,\lambda)}{q^2}\Phi(\vec k+\vec q,\eta)e^{-ik\eta}   
\end{eqnarray}
where, as usual, $\Phi(\vec q,\eta)=(2\pi)^{-3/2}\int d^3x e^{i\vec q \vec x} 
\Phi(\vec x, \eta)$.
The zeroth order equation implies $A_0^{(0)\vec k,\lambda}(\vec q,\eta)=0$.
The spatial equations (\ref{ieq}) can be written to first order as:
\begin{eqnarray}
2\Phi^\prime A_i^{(0)\prime} +\partial_i {A_0^{(1)\prime}}-{A_i^{(1)\prime\prime}}
+2\vec\nabla\Phi\cdot \vec\nabla
A_i^{(0)}-2\vec\nabla\Phi\cdot\partial_i \vec A^{(0)}+\vec \nabla^2 A_i^{(1)}
+4\Phi\vec \nabla^2 A_i^{(0)}=0
\label{speq}
\end{eqnarray}
Inserting again expansion (\ref{expansion}),  
these equations can be rewritten in 
Fourier space as:
\begin{eqnarray}
\frac{d^2}{d\eta^2}{A_i^{(1)\vec k, \lambda}}(\vec q,\eta)+q^2
A_i^{(1)\vec k, \lambda}(\vec q,\eta)-J_i^{\vec k, \lambda}(\vec q,\eta)=0
\label{EMeq}
\end{eqnarray}
where:
\begin{eqnarray}
J_i^{\vec k, \lambda}(\vec q,\eta)&=-&\sqrt{\frac{2k}{V}}\left(\left(i
\Phi'(\vec k+\vec q,\eta)
+\frac{k^2-\vec k \cdot \vec q}{k}\Phi(\vec k+\vec q,\eta)\right) 
\epsilon_i(\vec k,\lambda) e^{-ik\eta}\right. \nonumber \\
&+&\left.(\vec \epsilon(\vec k,\lambda)
\cdot \vec q) \;\Phi(\vec k+\vec q,\eta)\frac{k_i}{k}e^{-ik\eta}-i
\frac{\vec \epsilon(\vec k,\lambda)
\cdot \vec q}{q^2}\frac{d}{d\eta}\left(\Phi(\vec k+\vec q,\eta)
e^{-ik\eta}\right)
q_i\right)
\label{Jterm}
\end{eqnarray}

Solving these equations we find, up to first order in the perturbations:
\begin{eqnarray}
A_i^{\vec k, \lambda}(\vec q,\eta)=\frac{\epsilon_i
(\vec k,\lambda)}{\sqrt{2kV}}
\delta(\vec q-\vec k)e^{-ik\eta}+\frac{1}{q}\int_{\eta_0}^\eta 
J_i^{\vec k, \lambda}(\vec q,\eta')\sin(q(\eta-\eta'))d\eta'
\end{eqnarray}
where $\eta_0$ denotes the starting time of inflation. 
Comparing this expression  with (\ref{asin}), 
it is straightforward to obtain the Bogolyubov coefficients 
$\beta_{kq \lambda\lambda'}$,
they are given by:
\begin{eqnarray}
\beta_{kq \lambda\lambda'}=\frac{-i}{\sqrt{2qV}}\int_{\eta_0}^{\eta_1} 
\vec \epsilon \;(\vec q,\lambda')\cdot \vec J^{\;\vec k, \lambda}(\vec q,\eta) 
e^{-iq\eta}d\eta
\end{eqnarray}
where $\eta_1$ denotes the present time.  The total number of 
photons created with comoving wavenumber $k_G=2\pi/\lambda_G$, corresponding to
the relevant coherence length, is therefore given by \cite{Birrell}:
\begin{eqnarray}
N_{k_G}=\sum_{\lambda,\lambda'}\sum_k \vert \beta_{kk_G \lambda\lambda'}\vert^2
\label{total}
\end{eqnarray}
We will concentrate only in the effect of super-Hubble scalar perturbations 
whose
evolution is relatively simple \cite{brand}:
\begin{eqnarray}
\Phi(\vec k,\eta)=C_k\frac{1}{a}\frac{d}{d\eta}\left(\frac{1}{a}\int a^2
  d\eta\right)+ D_k\frac{a'}{a^3},
\label{perturb}
\end{eqnarray}
the second term decreases during inflation and can soon  be neglected. 
Thus, it will be useful to rewrite the perturbation 
as: $\Phi(\vec k,\eta)=C_k {\cal F}(\eta)$. 
During inflation or preheating, these perturbations evolve in time, 
whereas they are
practically constant during radiation or matter eras.
We will  neglect the effects of the perturbations once they  reenter the
horizon. This is a good approximation for modes reentering right after the end
of inflation since they are rapidly damped. In addition, we will show 
that those modes are
the more relevant ones in the calculation.

The power spectrum corresponding to (\ref{perturb}) is given by:
\begin{eqnarray}
{\cal P}_\Phi(k)=\frac{k^3 \vert C_k\vert ^2}{2\pi^2 V}=
A_S^2\left(\frac{k}{k_C}\right)^{n-1}
\label{power}
\end{eqnarray}
where for simplicity we have taken a power-law behaviour with spectral 
index $n$ and
we have set the normalization at the COBE scale 
$\lambda_C\simeq 3000 \;\mbox{Mpc}$ 
with $A_S\simeq 5 \cdot 10^{-5}$. In the case of a blue spectrum, 
with positive tilt ($n>1$),
perturbations will grow at small scales and it is necessary to introduce a 
cut-off $k_{max}$ in order to avoid
excessive primordial black hole production \cite{kolb}. Accordingly, 
only below the
cut-off the perturbative method will be reliable. For negative tilt or 
scale-invariant
spectrum there will be also a small scale cut-off related to the minimum size
of the horizon  $k_{max}\lsim a_I H_I$, where the $I$ subscript
denotes the end of inflation.

We can obtain an explicit expression for the total number of photons 
(\ref{total})  
in terms of the power spectrum. Taking the continuum limit 
$\sum_k\rightarrow (2\pi)^{-3/2}V\int d^3k$, we get:
\begin{eqnarray}
N_{k_G}&=&\sum_{\lambda,\lambda'}V\int \frac{d^3k}{(2\pi)^{3/2}}\vert 
\beta_{kk_G\lambda\lambda'}\vert^2\nonumber \\
&=&\sum_{\lambda,\lambda'}V\int \frac{d^3k}{(2\pi)^{3/2}}
\frac{\vert C_{\vert k +k_G \vert}\vert^2}{2k_GV^2}\left\vert\int d\eta \left(
\sqrt{2k}\left(\left(i
{\cal F}\;'
+\frac{k^2-\vec k \cdot \vec k_G}{k}{\cal F}\right) 
(\vec\epsilon(\vec k,\lambda)\cdot\vec\epsilon(\vec k_G,\lambda'))
\right.\right.\right.\nonumber \\
&+&\left.\left.\left.(\vec \epsilon(\vec k,\lambda)
\cdot \vec k_G) \;(\vec\epsilon(\vec k_G,\lambda')
\cdot \vec k) \;\frac{\cal F}{k}\right)e^{-i(k_G+k)\eta}\right)
\right\vert^2
\end{eqnarray}
Notice that the last term
in (\ref{Jterm}) does not contribute to $\beta_{kq \lambda\lambda'}$
because of the transversality condition of the polarization vectors.
The integration in $d^3k$ is dominated by the upper limit, i.e. 
$k\gg k_G$ and accordingly
we can ignore the effect of the terms proportional to  $\vec k_G$.
In addition, for those modes $k$ which are outside the Hubble
radius
at the end of inflation, we have
 $k\eta\ll 1$. With these simplifications we obtain:
\begin{eqnarray}
N_{k_G}&\simeq&
\sum_{\lambda,\lambda'}\int \frac{dk\;d\Omega}{(2\pi)^{3/2}}
\frac{\vert C_k \vert^2k^2}{2k_GV}\left\vert\int d\eta \left(
\sqrt{2k}\left(\left(i
{\cal F}\;'
+k {\cal F}\right) 
(\vec\epsilon(\vec k,\lambda)\cdot\vec\epsilon(\vec k_G,\lambda'))\right)
\right)\right\vert^2
\end{eqnarray} 
Performing the integration in the angular variables and using the definition
of the power spectrum in (\ref{power}), we obtain:
\begin{eqnarray}
N_{k_G}&\simeq&\frac{4(2\pi)^{3/2}}{3k_G}\int dk A_S^2
\left(\frac{k}{k_C}\right)^{n-1}
\left\vert\int d\eta \left(i
{\cal F}\;'
+k {\cal F}
\right)\right\vert^2
\end{eqnarray} 
Finally, we will estimate the time integral. The behaviour of 
scales that reenter the
horizon during the radiation dominated era is oscillatory  with a decaying 
amplitude \cite{brand}, therefore, there is no long-time contribution to the 
integral that could spoil the perturbative method.
Thus, for simplicity we will assume that the
function ${\cal F}$ vanishes  for $\eta \geq 1/k$, and accordingly 
we estimate, 
$\left\vert\int d\eta \left(i {\cal F}\;' +k {\cal F}
\right)\right\vert^2\sim \Od(1)$. Our final expression for the 
occupation number
is:
\begin{eqnarray}
N_{k_G}&\simeq&\frac{4(2\pi)^{3/2}A_S^2}{3k_G(k_C)^{n-1}}\int_{k_C}^{k_{max}} 
dk k^{n-1}\simeq
\frac{4(2\pi)^{3/2}A_S^2 }{3\,n}\frac{k_{max}^n}{k_G\;k_C^{n-1}}
\label{final}
\end{eqnarray} 

\section{Magnetic field generation}

The  energy density stored in a magnetic field
mode $B_k$ with wavenumber $k$ is given by:
\begin{eqnarray}
\rho_B(\omega)=\omega \frac{d\rho_B}{d\omega}=\frac{\vert B_k
  \vert^2}{2},
\label{magn}
\end{eqnarray}
with $\omega=k/a$ the physical wavenumber. In terms of the occupation number, 
it reads;
\begin{eqnarray}
\rho_B(\omega)=\omega^4 N_k.
\label{bogen}
\end{eqnarray}
From (\ref{final}), we can obtain the  strength of the  field at
decoupling on a coherence scale corresponding to
$k_G\sim  10^{-36}$ GeV as:
\begin{eqnarray}
\vert B_{k_G}^{dec}\vert \simeq \sqrt{2}(\omega_G^{dec})^2\;
N_{k_G}^{1/2}\simeq
\frac{2^{3/2}(2\pi)^{3/4}\;A_S}{\sqrt{3n} \;a_{dec}^2}\;\frac{k_{max}^{n/2}\; 
k_G^{3/2}}{k_C^{(n-1)/2}}
\label{mag}
\end{eqnarray}

In Fig.1 we have plotted the strength of the magnetic field
generated as a function of the comoving cut-off frequency 
$k_{max}$ for different
values of the spectral index $n$. Notice
that  the results are in
general too weak to explain the observed fields without any
amplification. However, for certain values of the cosmological 
parameters, 
the produced fields
could act as seeds for a galactic dynamo.
\begin{figure}
\vspace{-4cm}
\begin{center}
\mbox{\epsfysize=13.5cm\epsfxsize=13.5cm
\epsffile{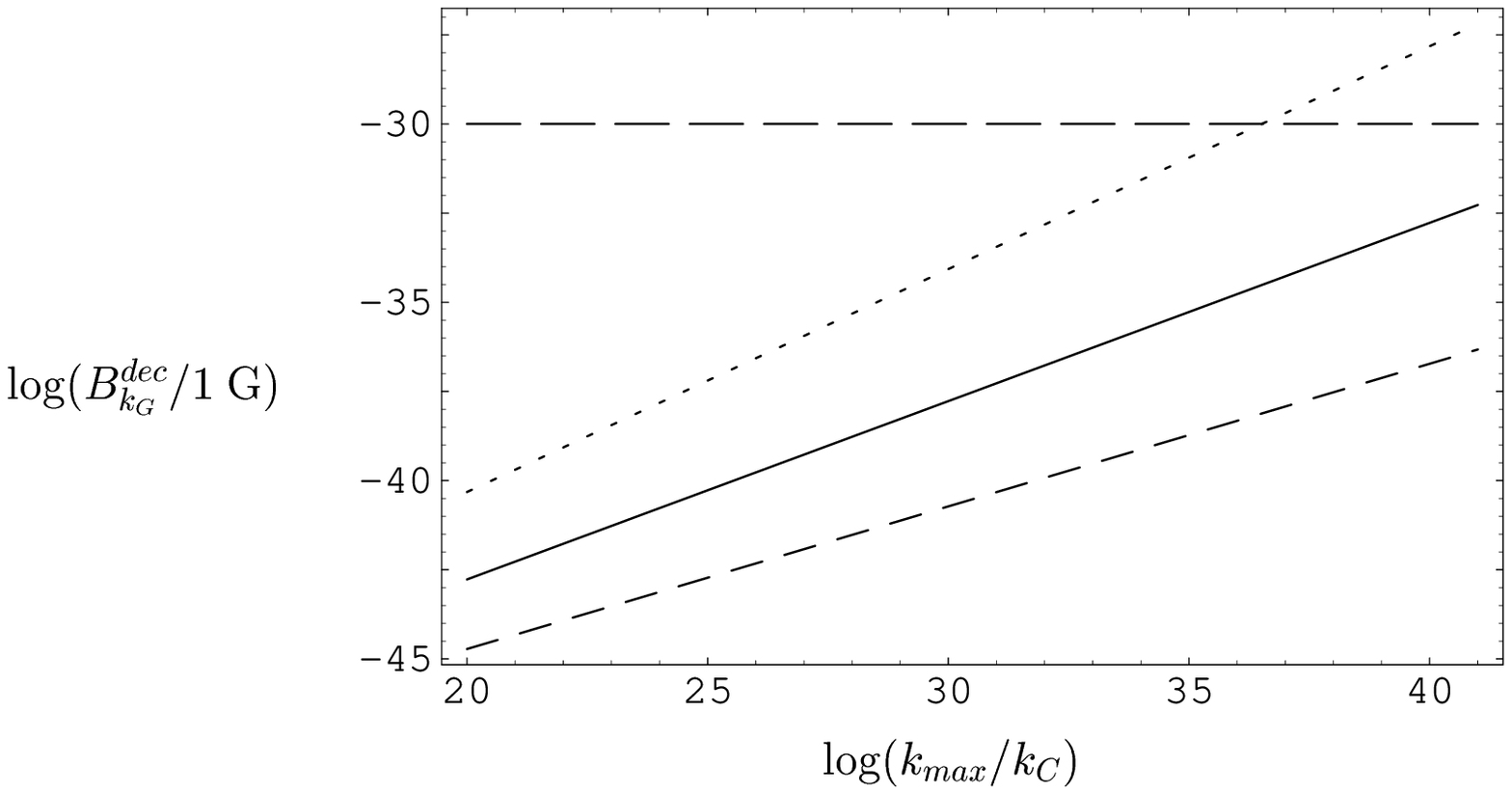}}
\end{center}
\leftskip 1cm
\rightskip 1cm
{\footnotesize
{\bf Figure 1.-} $\log(B_{k_G}^{dec}/1\;\mbox{G})$ as a function of
$\log(k_{max}/k_C)$. The continuous line corresponds to the scale-invariant
Harrison-Zeldovich spectrum with  
$n=1$,
the dashed line to $n=0.8$  and the dotted line to
$n=1.25$. The dashed horizontal line represents the weakest galactic dynamo 
seed field limit corresponding to a 
flat universe with cosmological constant and 
$h=0.5$.}
\end{figure}

We see that the spectrum of
magnetic
fields produced by this mechanism is  thermal $B_k\sim k^{3/2}$,
 in the low-momentum region. We can then compare this spectrum
with that corresponding to the thermal background radiation with a temperature
$T_{dec}\simeq 0.26$ eV present at decoupling time. The energy density 
in photons with comoving wavenumer $k_G$ at decoupling 
is given by $\rho_R(\omega_G) \simeq k_G^3 T_{dec}/a_{dec}^3$. Thus we find:
\begin{eqnarray}
\frac{\rho_R(\omega_G)}{\rho_B(\omega_G)}=\frac{a_{dec} T_{dec}}{N_{k_G}k_G}
\simeq 
1.4 \cdot 10^{36} \left(\frac{k_C}{k_{max}}\right)^n
\end{eqnarray}
From this expression we see that the magnetic field energy
density will dominate over the
background thermal radiation whenever $\log(k_{max}/k_C)\gsim 36/n$, i.e.
for example, for $n=1$ this implies $k_{max} \gsim  10^{-6}$ GeV. 

 The cut-off frequency
 $k_{max}$ cannot  be easily determined in general, since
it  depends on the specific mechanism that generates the perturbations and 
also on the 
evolution of the universe during  reheating and thermalization.
However we can
estimate typical values in some particular regime. In the case in which 
metric perturbations are generated by inflation, it is natural to expect, 
$k_{max}\lsim a_IH_I$, as commented before.
 Thus, let us take the simplest chaotic inflation model with potential 
$V(\phi)=\lambda \phi^4/4$ \cite{Linde}, 
with $\lambda\simeq 10^{-12}$ fixed by COBE. In this model
the Hubble parameter during inflation is $H_I\simeq 10^{13}$ GeV.
Owing to the uncertainties
commented before, we will let the reheating temperature $T_{RH}$ be 
a free parameter.
After reheating the universe evolution is adiabatic 
$a_I/a_{dec}\sim T_{dec}/T_{RH}$, and we can calculate the
cutoff frequency as $k_{max}/k_C\sim a_IH_I/k_C\sim 
a_{dec}T_{dec}H_I/(T_{RH}k_C)$, which yields 
$k_{max}/k_C\sim 10^{42}$ GeV/$T_{RH}$. 
 Here we have  assumed that the
inflation-radiation transition takes place
in a few inflaton oscillations \cite{Linde} (see also
\cite{Linde2}).
Comparing with Fig. 1 we see that
with these simple estimations for the  $\lambda\phi^4/4$ model,
the amplification
could be above the requirements
of the galactic dynamo if $T_{RH}\lsim 10^6$ GeV for $n=1.25$. 

\section{Conductivity effects}

In the previous discussion we have assumed that electric conductivity
of the universe played no role in the generation of the magnetic
fields. Although neglecting conductivity is a good approximation 
during inflation, it is not
during the reheating or radiation periods. As commented
before, the copious production
of particles during reheating produces the growing of the conductivity
which becomes very high during radiation \cite{turner}. This implies that
the magnetic fields produced in the inflation-radiation transition will
evolve conserving magnetic flux  $\rho_B\sim a^{-4}$.
However, it has been recently showed \cite{Gio} that the growth of conductivity 
during reheating could affect
the evolution of the EM modes.
The effects of conductivity can be taken into account in a phenomenological
way by introducing a current source $J_i=\sigma_c a A_i'$ in
(\ref{speq}). This approach is only valid at sufficiently large scales 
\cite{Gio} and in general the rigorous treatment  would requiere to solve
the set of coupled EM-matter fields equations (Vlasov equations) which
is
beyond the scope of this paper.
In general, $\sigma_c(\vec x, \eta)$ is a time and
space dependent function, and it has been shown that it  grows exponentially
during parametric resonance in a non-equilibrium plasma in QED \cite{Boy}. 
The calculation of the actual function 
$\sigma_c(\vec x, \eta)$ in our case would be rather involved and model
dependent,
since it would require information about the reheating and
thermalization processes. 
For that reason we will not take any particular model, but
we will do a general discussion of the possible effects 
in different
cases. We will also
assume for simplicity that all the Fourier modes of the conductivity have
the same time evolution, i.e. $\sigma_c(\vec k,\eta)=\Sigma(\eta)\sigma_k$
during reheating.

In previous works \cite{calz,Gio}, the conductivity was considered as
an homogeneous field, $\sigma_c(\eta)$ and because of the form of the
current source term, its effect was the damping of each EM mode.
In our production mechanism, the inhomogeneities are able
to mix different EM modes and for this reason we
need to have information about the complete spectrum $\sigma_k$
and not only about the large-scale components. In addition, our
analysis is perturbative and therefore we can only describe
the initial stages in which the conductivity is still small. 
Let us then consider the modified spatial equation to 
first order in the metric perturbations and the conductivity:
\begin{eqnarray}
(2\Phi^\prime -a\sigma_c)A_i^{(0)\prime} 
+\partial_i {A_0^{(1)\prime}}-{A_i^{(1)\prime\prime}}
+2\vec\nabla\Phi\cdot \vec\nabla
A_i^{(0)}-2\vec\nabla\Phi\cdot\partial_i \vec A^{(0)}+\vec \nabla^2 A_i^{(1)}
+4\Phi\vec \nabla^2 A_i^{(0)}=0
\end{eqnarray}
Following a similar analysis we find  the same equation (\ref{EMeq}),
but with the new current:
\begin{eqnarray}
J_i^{\vec k, \lambda}(\vec q,\eta)&=-&\sqrt{\frac{2k}{V}}\left(\left(i
\Phi'(\vec k+\vec q,\eta)-\frac{i}{2}a\;\sigma_c(\vec k+\vec q,\eta)  
+\frac{k^2-\vec k \cdot \vec q}{k}\Phi(\vec k+\vec q,\eta)\right) 
\epsilon_i(\vec k,\lambda)e^{-ik\eta}\right. \nonumber \\
&+&\left.(\vec \epsilon(\vec k,\lambda)
\cdot \vec q) 
\;\Phi(\vec k+\vec q,\eta)\frac{k_i}{k}e^{-ik\eta}
-i\frac{\vec \epsilon(\vec k,\lambda)
\cdot \vec q}{q^2}\frac{d}{d\eta}
\left(\Phi(\vec k+\vec q,\eta)e^{-ik\eta}\right)
q_i\right)
\end{eqnarray}
In \cite{Gio} the following  lower limit on the (homogeneous) conductivity 
is obtained $a_I\,\sigma_c \sim
a_I\,H_I/\alpha$, with $\alpha$ the fine structure constant, i.e. 
$a_I\,\sigma_c \sim k_{max}/\alpha$ and this implies that the conductivity
term will dominate in $J_i^{\vec k, \lambda}(\vec k_G,\eta)$
for $k\ll k_{max}$. However, as commented before, the dominant contribution
to the EM amplification  comes
from the high-frequency modes, i.e. 
$k\sim k_{max}$. In such case, the importance of the conductivity term
is determined by the ratio $\sigma_k/C_k$ when 
$k\rightarrow k_{max}$. 
Thus, if the conductivity is almost homogeneous, 
its spectrum will decline at short scales  and 
we expect its contribution to the EM field evolution to
be negligible. In the opposite case, the analysis would become much more
involved, since the above magnetohydrodynamical approximation would break down
and the full set of microscopic equations would be needed.
In any case, this simple analysis shows that the possible damping effects 
mainly affect those modes with $k \sim k_G$ and that, depending on the actual
conductivity spectrum, the rest of modes could be 
 less severely affected than in other models.

\section{Conclusions}

In this work we have studied the production of photons in the presence
of an inhomogeneous gravitational background. We have shown how the
breaking of conformal invariance induced by the
evolution of metric perturbations in the inflation-radiation
transition is able to produce particles, and
we have related the occupation number with the scalar metric perturbations 
power spectrum. 

We have considered the possibility that this mechanism could have had
some relevance in the problem of  galactic magnetic fields and
we have concluded that the total amplification is several orders of magnitude
below the observed strengths. However, for certain values of the
cosmological parameters and with the assistance of the dynamo mechanism,
the amplification could be compatible with the current (low-redshift) galactic
observations. We have also considered the effect of conductivity in
a phenomenological way and show that although it could affect the
evolution of EM modes, in some cases and depending on the particular 
form of the spectrum, the effects could be small.

The mechanism studied in this work only 
relies on the existence of a primordial spectrum of metric perturbations, 
described by the scalar spectral index $n$ and a possible cutoff frequency 
$k_{max}$, which are the only free parameters in the model.
Therefore, as a by-product we get that magnetic fields  could also   
provide useful information about  the  metric 
perturbations spectrum and in particular about its small-scales 
region.

{\bf Acknowledgements:} This work
has been partially supported by the Ministerio de Educaci\'on y
Ciencia (Spain) (CICYT AEN 97-1693).

\end{document}